\documentclass{article}

\usepackage[main,final]{neurips_2026}

\usepackage{amsmath}
\usepackage{amssymb}
\usepackage{mathtools}
\usepackage{amsthm}
\usepackage{multirow}
\usepackage{multicol}
\usepackage{tikz}
\usepackage{enumitem}
\DeclareMathOperator*{\argmin}{argmin}

\usepackage{algorithm}
\usepackage{algorithmic}
\usepackage{graphicx}
\usepackage{wrapfig}
\usepackage{float}

\usepackage[utf8]{inputenc} 
\usepackage[T1]{fontenc}    
\usepackage{hyperref}       
\usepackage{url}            
\usepackage{booktabs}       
\usepackage{amsfonts}       
\usepackage{nicefrac}       
\usepackage{microtype}      
\usepackage{xcolor}         

\title{vExpert: Virtualizing Expert Storage for Adaptive Load Balancing in Distributed MoE Inference}

\author{%
  Wenxun Wang\textsuperscript{1,3}
  \quad
  Xiuhong Li\textsuperscript{2}
  \quad
  Yida Wang\textsuperscript{2}
  \quad
  Chen Tang\textsuperscript{1}
  \quad
  Zongle Huang\textsuperscript{1,3}
  \quad
  Ke Hong\textsuperscript{1}
  \\[0.2ex]
  \textbf{Yu Wang}\textsuperscript{1}
  \quad
  \textbf{Yongpan Liu}\textsuperscript{1,3\thanks{Corresponding Author.}}
  \\[0.3ex]
  \small
  \textsuperscript{1}Tsinghua University
  \quad
  \textsuperscript{2}Infinigence-AI
  \quad
  \textsuperscript{3}BNRist
  \\[-0.1ex]
  \small
  \texttt{\{wx-wang23\}@mails.tsinghua.edu.cn}
  \quad
  \texttt{\{ypliu\}@tsinghua.edu.cn}
}

\begin{document}

\maketitle

\begin{abstract}
Mixture-of-Experts (MoE) models have become prevailing, yet their distributed deployment suffers from severe dynamic load imbalance due to the conflict between static expert placement and inherent routing dynamism.
While host DRAM offers capacity to avoid static device storage for load balancing, we identify existing offloading works fundamentally fail in distributed inference due to two overlooked challenges: (1) coupled performance impacts of runtime Host-to-Device transfers due to load dynamism and shared PCIe topology; (2) system-level memory overhead incurred by DRAM usage at scale.
In this paper, we propose \textsc{vExpert} that shifts DRAM offloading paradigm from partial residency to full redundancy through virtualized expert storage. 
We pioneer the first systematic performance model that jointly captures H2D latency, PCIe contention, and load imbalance, building a lightweight allocator to dynamically map physical experts to virtual slots runtime.
A novel disaggregated manager is designed to reduce memory overhead in multi-process environments.
Evaluated on DeepSeek-V3.1 within an 8-node H100 cluster, \textsc{vExpert} improves load balance ratio by up to 55\% and achieves 1.2$\times$ end-to-end speedup. 
\end{abstract}

\section{Introduction}
The scaling of Large Language Models (LLMs) \cite{llm1,gpt4} has established Mixture-of-Experts (MoE) \cite{moe0,moe1,glm} as the \textit{de facto} architecture for efficient inference. To handle massive model sizes, Expert 
\begin{wrapfigure}{r}{0.5\textwidth}
\vspace{-0.7cm}
    \vskip 0.2in
    \centering
\includegraphics[width=\linewidth,height=0.28125\textwidth]{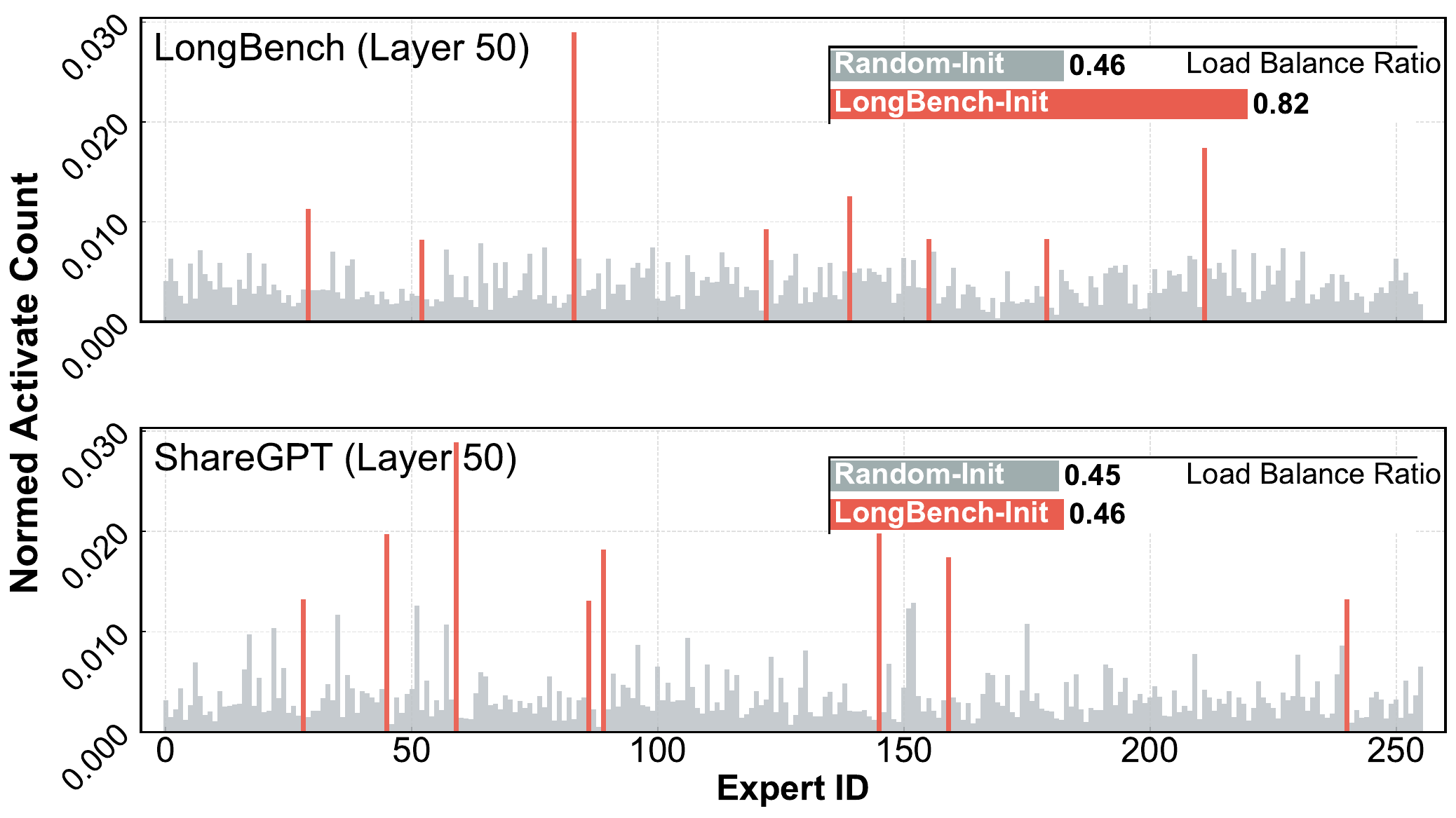}
    \caption{Distribution of activated expert for DeepSeek-V3.1. Random-init and LongBench-init denote expert placement initialized randomly and via LongBench statistics, respectively. }
    \label{fig_load_imbalance}
\vspace{-0.5cm}
\end{wrapfigure}
 Parallelism (EP) is universally adopted for distributed inference, sharding experts across multiple GPUs \cite{vllm,sglang,ep1}.

However, the static expert placement contrasts with the inherent dynamism of MoE \cite{tutel,omni-infer}. As shown in \autoref{fig_load_imbalance}, activated experts fluctuates drastically depending on the input \cite{eaas}, leading to severe load imbalance where GPUs hosting hot experts become bottlenecks. 
Existing systems mitigate this by replicating redundant experts \cite{lina,deepseekv3,deepspeed-moe}. 
Yet, this approach exposes a fatal flaw: \textbf{the inherent conflict between redundancy capacity and memory constraints.} Redundancy capacity can be viewed as the ratio between the number of experts visible to a GPU and the total number of experts. While GPU HBM is heavily occupied by model parameters and KV caches, it restricts redundancy to merely 1--2 experts per GPU, stripping the system of the elasticity required to handle rapid load drifts.

We identify a \textit{tiered expert storage hierarchy} as the key to decouple redundancy capacity from HBM limits.
This insight is motivated by three observations in modern AI servers:
First, the underutilized TB-scale host DRAM provides an abundant capacity for full-model redundancy.
Second, the aggregated Host-to-Device (H2D) unidirectional bandwidth (e.g., 512GB/s via PCIe Gen5) is competitive with Device-to-Device interconnects (e.g., 450GB/s via NVLink 4.0) \cite{beluga,tasp}, rendering runtime weight streaming viable. 
Third, host memory offers topology symmetry and global visibility, enabling uniform access by all intra-node GPUs.
Together, these factors motivate treating host DRAM as a second-level storage to break the coupling between redundancy capacity and device HBM.

\begin{figure}[!t]
  \vspace{-0.5cm}
  \vskip 0.2in
  \begin{center}
\centerline{\includegraphics[width=\textwidth,height=0.25\textwidth]{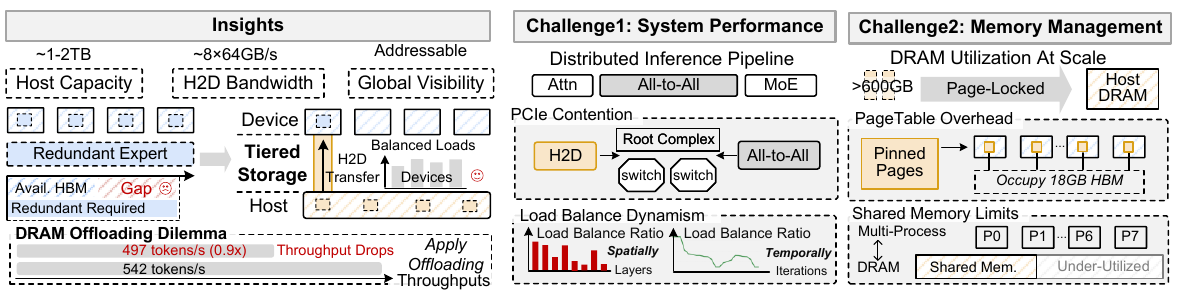}}
    \caption{
      Insights and challenges of utilizing host resources for distributed inference at scale. 
    }
    \label{fig2}
  \end{center}
  \vspace{-1.0cm}
\end{figure}

A straightforward approach to exploiting host resources is DRAM offloading \cite{es-moe,moe-lightning,swapmoe}, which adopts a partial residency where a subset of experts resides in HBM. It fetches GPU-missed experts runtime and manages H2D transfers based on overlapping, assuming that H2D transfers incur negligible overhead when masked by concurrent computation.
However, applying offloading for distributed inference incurs fundamental challenges as shown in \autoref{fig2}.

First, the impact of H2D transfers on \textbf{system performance} is inherently complex and structurally coupled. On one hand, distributed MoE inference relies on inter-node All-to-All (A2A) communication. Integrating H2D transfers into inference pipeline causes contention for node-level PCIe resources, leading to negative latency impacts.
On the other hand, load balance exhibits dynamism both spatially (layers) and temporally (iterations). 
DRAM offloading fundamentally fails as it incurs constant PCIe traffic for GPU-miss regardless of actual load distribution and PCIe contention, underscoring the necessity to model the coupled impact on system performance.

Second, leveraging host DRAM at scale incurs \textbf{system-level memory overheads} in multi-process servers. Achieving peak H2D bandwidth requires page-locked weights. As each GPU process must maintain its own page table mappings, serving a 671B model consumes GBs of HBM, limiting KV cache capacity—which in turn constrains concurrency and degrades throughput.
Besides, exposing the host-resident experts across multiple processes relies on OS-level shared memory, whose default quotas are strictly capped. It reduces the accessible host capacity and prevents full-model redundancy.

To address these challenges, we present \textsc{vExpert}, a virtualized expert storage system for MoE decoding.
It proposes a paradigm shift from partial residency to full redundancy by storage abstraction, virtual Expert (vExpert), that decouples expert residency across a tiered Device-HBM and Host-DRAM hierarchy.
To guarantee system performance, we pioneer the first systematic analysis of H2D transfers in distributed inference and build a fine-grained performance model that quantifies the coupling impact of H2D latency, PCIe contention, and load imbalance.
Guided by this model, we design a lightweight allocator that dynamically remaps physical experts to virtual slots in response to shifting workloads across layers and iterations.
To mitigate system-level memory overheads, we implement a novel memory management that decouples data provisioning (coordinator) from consumption (worker). It reduces OS-level addressing overhead and bypasses shared memory limits. 

We implement \textsc{vExpert} and evaluate it on an 8-node H100 cluster.
Compared to SOTA baselines, \textsc{vExpert} improves load balance by up to 55\% under dynamic workloads, translating to up to 20\% higher end-to-end decoding throughput.
These results demonstrate that \textsc{vExpert} is a practical and effective path toward elastic and robust MoE inference. Code is available at \href{https://github.com/Flying-Cloud/vExpert}{https://github.com/Flying-Cloud/vExpert}.

\section{Related Work}
\subsection{MoE Serving At Scale}

\begin{wrapfigure}{r}{0.6\textwidth}
\vspace{-1.5cm}
    \centering
\includegraphics[width=\linewidth,height=0.19\textwidth]{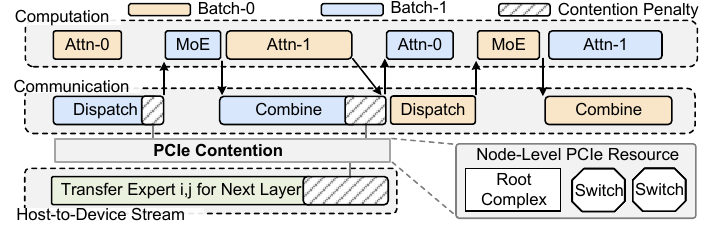}
    \caption{Distributed inference timeline with H2D transfers.}
    \label{fig_timeline}
\vspace{-0.2cm}
\end{wrapfigure}

\textbf{Inference Timeline.} MoE layers follow a structured pipeline of four stages in distributed inference \cite{sglang}: (1) \texttt{Attention}, (2) \texttt{Dispatch} that scatters tokens to the selected experts, (3) \texttt{MoE} that computes the tokens through its local experts, (4) \texttt{Combine} that gathers the processed tokens back to their original GPUs. To hide the overhead of communication, two-batch-overlap (TBO) is commonly employed to enhance throughput \cite{deepseekv3} as shown in \autoref{fig_timeline}.

\textbf{Load Imbalance.}
The critical path is dominated by MoE and All-to-All. Load imbalance directly impacts the latency of these operations and greatly affects system throughput \cite{load_balance1}.
Maintaining workload balance in distributed system is crucial yet challenging \cite{step3}. 
An expert placement that is well balanced for one workload can quickly become imbalanced under load drift, rendering current static or offline optimization insufficient. 
This gap necessitates a system that can adapt to dynamic load shifts which is commonly observed in online serving.

\subsection{DRAM Offloading}
DRAM offloading has been explored to utilize host resources for efficient LLM inference \cite{moe_opt,floe,fiddler,rapidmoe,fluxmoe}. 
These systems typically adopt a partial residency paradigm \cite{moe-lightning,flexgen}, partitioning weights between GPU HBM and host DRAM, where H2D transfers are triggered by GPU misses. 
To minimize overhead, they rely on overlapping that assumes H2D transfers incurs zero overhead if masked by concurrent computation. While viable for single-node or single-device setups, it fails in distributed inference as H2D transfers compete for limited node-level PCIe resources with \texttt{Dispatch} and \texttt{Combine}. It exacerbates physical contention on the critical path, making constant H2D transfer sub-optimal for system performance.
\cite{ktransformers} furher offloads both storage and computation to CPUs, exploiting available CPU compute resources to meet latency requirements. Though suitable for small-batch edge serving, the limited CPU capabilities cannot sustain the strict latency demands of online serving under massive concurrency (e.g., batch size > 100). Consequently, DRAM offloading has primarily been restricted to the latency-tolerant prefillling or training \cite{es-moe,mooncake,pipeoffload}. 
Our work pioneeringly leverages host resources for distributed MoE decoding, revealing and addressing challenges in system performance and memory management.

\section{\textsc{vExpert} Paradigm}
\label{sec3}
\subsection{System Architecture}
\begin{figure*}[!t]
  \vskip 0.2in
  \begin{center}
\centerline{\includegraphics[width=1.0\textwidth,height=0.36\textwidth]{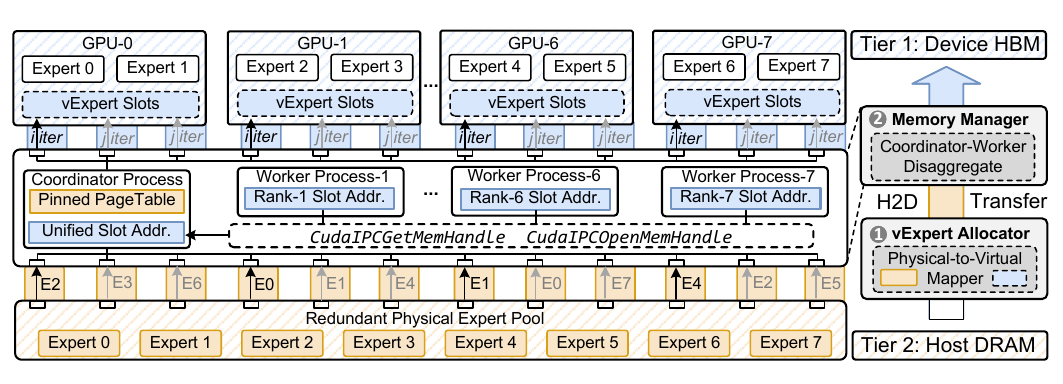}}
    \caption{
      Overview of the proposed \textsc{vExpert} storage hierarchy and system design.
    }
    \label{fig_overview}
  \end{center}
\vspace{-1.0cm}
\end{figure*}

\textbf{Overview.} We introduce vExpert abstraction to decouple expert residency from GPU memory constraints. vExperts are logically visible to GPUs as locally available experts but physically reside in host DRAM.
In this paradigm, the entire expert pool is exposed and addressable to every GPU process, thereby enabling non-blocking, dynamic configuration of redundant experts to adapt to shifting workloads.
GPUs maintain a minimal set of vExpert Slots—HBM buffers that serve as destination targets for H2D transfers. The \textsc{vExpert} architecture establishes a tiered expert storage hierarchy (\autoref{fig_overview}): 

\textbf{L1:Device HBM.} L1 does not store static expert replicas. Instead, it serves as a dynamic cache for weights required by the current iteration and layer.

\textbf{L2:Host DRAM.}  Leveraging the high capacity of DRAM, L2 maintains a full-coverage physical expert pool in the model.

\textbf{Pre-Allocating Pipeline.} 
 vExperts are managed via a layer-wise pre-allocating pipeline to mask the H2D latency in tiered hierarchy. During the execution of layer $i$, system asynchronously initiates H2D transfers for the vExperts required by layer $i+1$. By overlapping H2D transfer with the computation of the preceding layer, vExperts are available in their designated slots just-in-time for execution. As a result, GPUs only need to maintain buffers sufficient to hold the virtual experts of two layers (current and next).

\textbf{vExpert Allocator.} 
The system utilizes the allocator to dynamically resolve the optimal mapping between L2 physical weights and L1 virtual slots, specifying the per-layer vExpert number and the physical expert ID assigned to each GPU.
\autoref{fig_overview} shows an example: in iteration i, each device hold a single vExpert and slot of GPU-0 is mapped to physical expert E2. In later iteration j, allocator configures two vExperts per device, with GPU-0 mapping to E3 and E6.

\textbf{Memory Manager.}
The manager implements a disaggregated scheme. A single coordinator process registers the global pinned memory for L2 physical pool and orchestrates H2D transfers. Worker processes allocate their local vExpert slot buffers at L1, which are made visible to coordinator via CUDA Inter-Process Communication (IPC).
It allows coordinator to perform transfers into slots of any workers.
Details of each component are presented in Section \ref{sec:sys}.

\subsection{Performance Modeling and Optimization}
\label{sec:3.2}
We develop a principled performance model that quantifies the impacts of vExperts.
\subsubsection{Latency Model Formulation}
The latency of MoE layer is typically dominated by three sequential operations: \texttt{Dispatch}, \texttt{MoE}, and \texttt{Combine}. Their aggregate duration become the primary determinant of overall latency in the pipeline. We therefore define the baseline per-layer latency as:
\begin{equation}
    T_i(L_i) = T_{i,\text{dispatch}}(L_i) + T_{i,\text{moe}}(L_i) + T_{i,\text{combine}}(L_i)
\end{equation}
$L_i = \frac{\text{AvgTokensPerGPU}}{\text{MaxTokensPerGPU}} \in [0,1]$ represents the load balance ratio, which determines the token number $B_{\max} = B / L_i$ in the most heavily loaded GPU ($B$ is the batch size). 
Both communication and computation are bottlenecked by $B_{\max}$ in the cluster. 

\textbf{Extension with Virtual Experts.} Introducing $V$ virtual experts per layer modifies the latency by: (1) improved load balance ratio ($L_i$), and (2) injecting H2D transfer overhead. To simplify the analysis without loss of generality, we aggregate the \texttt{Dispatch} and \texttt{Combine} into a single term $T_{i,\text{a2a}}$, the enhanced latency model becomes:

\vspace{-0.4cm}
\begin{equation}
\small
\begin{split}
        T_{i}^*(L_i, V) = T_{i,\text{a2a}}(L_i) + T_{i,\text{moe}}(L_i) 
    - \underbrace{T_{i,\text{a2a}}\cdot \gamma_1(L_i,V)}_{\text{communication gain}} - \underbrace{T_{i,\text{moe}}\cdot \gamma_2(L_i,V)}_{\text{computation gain}} + \underbrace{T_{\text{H2D}}(V)\cdot \eta}_{\text{H2D overhead}}
\end{split}
\label{eq:layer_time_model}
\end{equation}
$\gamma_1$ and $\gamma_2$ represent the latency reduction ratios. $T_{\text{H2D}}(V)$ is the latency to transfer $V$ experts, and $\eta$ captures the negative impact of PCIe contention. We proceed to analyze and profile each component based on setup in Section \ref{sec:5.1}.

\subsubsection{Positive Impact: Load Balancing Gains}
\label{sec:3.2.1}
\textbf{All-to-All Communication Gain ($\gamma_1$).} 
The latency of All-to-All scales with the maximum token number $B_{\max}$ handled by any rank \cite{lina}. A higher balance ratio $L$ decreases $B_{\max}$, leading to the reduction of latency as:
\begin{equation}
\small
\label{eq:gamma1}
\gamma_1(L_i, V) \approx 1-\frac{B/(L_i+\Delta_L)}{B/L_i} = \frac{\Delta(L_i,V)}{L_i + \Delta(L_i,V)}
\end{equation}
$\Delta(L_i, V)$ denotes the improvements in load balance.

\textbf{MoE Computation Gain ($\gamma_2$).} $L_i$ directly affects the \text{TokensPerExpert} (TPE, referred to $B_{\max}$) for MoE GEMM kernels.
The computation gain of MoE computation exhibits workload-dependent behavior \cite{sarathi-serve}:
\textbf{(1) Compute-Bound:} For large TPE ($\text{TPE} \geq 160$ in our profiling), MoE GEMMs are compute-bound, and latency scales proportionally with the maximum token count. In this regime, $\gamma_2\approx \gamma_1$.
\textbf{(2) Memory-Bound:}  For small TPE, MoE GEMMs are memory-bound where memory access overheads dominate. Reducing the token count yields diminishing returns as $\gamma_2 \approx 0$. 
\begin{wrapfigure}{r}{0.5\textwidth}
\vspace{-0.9cm}
    \vskip 0.2in
    \centering
\includegraphics[width=\linewidth,height=0.25\textwidth]{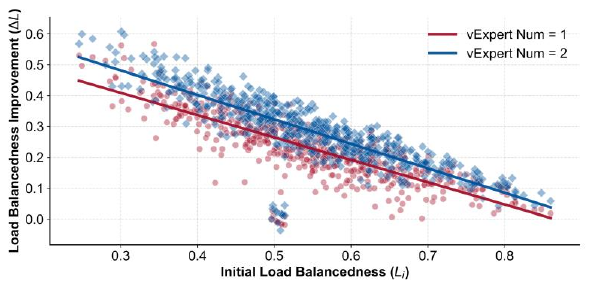}
    \caption{Analysis of load balance improvement ($\Delta$) through virtual experts on DeepSeek-V3.1.}
    \label{fig_delta}
\vspace{-0.8cm}
\end{wrapfigure}

We provide detailed profiling for these operations in Appendix \ref{apx:profiling_balance} for validation.

\textbf{Load Balance Improvement ($\Delta$).} $\Delta(L_i, V)$ is a complex function of: distribution of activated experts, initial load imbalance $L_i$, and vExpert number $V$. Closed-form modeling is intractable due to the stochastic nature of online request distributions. Instead, we characterize this relationship through offline profiling on representative datasets. Our observations indicate that for a fixed $V$, $\Delta$ can be approximated via a linear function as shown in \autoref{fig_delta}. Therefore, we approximate $\Delta$ via a first-order linear model:
\begin{equation}
\Delta(L_i, V) = \alpha\cdot L_i + \beta \cdot V + k
\end{equation}
Profiled on representative datasets, the fitted parameters are $\alpha,k,\beta = -0.75,0.1,0.53$. This empirical characterization captures load balancing effects for each iteration and layer.

\subsubsection{Negative Impact: H2D Transfer Overhead}
\label{sec:3.2.3}
\textbf{Path Topology and Interference.}
An non-intuitive challenge is the physical resource contention between H2D transfer and All-to-All communication \cite{deepseek-isca}. All-to-All typically traverses the GPU $\leftrightarrow$ PCIe Switch $\leftrightarrow$ NIC $\leftrightarrow$ PCIe Switch $\leftrightarrow$ GPU. H2D transfer traverses the CPU $\rightarrow$ PCIe Switch $\rightarrow$ GPU. While PCIe supports full-duplex, both traffic flows compete for the resources of PCIe Switch and Root Complex. We model this interference via an additive penalty $T_{\text{H2D}}(V) \cdot \eta$, where $T_{\text{H2D}}$ is both the time of H2D transfer and the duration of bandwidth contention, $\eta$ reflects the contention loss for All-to-All. We derive $\eta$ and $T_{\text{H2D}}$ via benchmarking concurrent bandwidth in Appendix \ref{apx:profiling_pcie}.

\textbf{H2D Latency ($T_{\text{H2D}}$).}  Transferring $V$ experts from Host to Device costs $T_{\text{H2D}}(V) = \frac{V \cdot M_{\text{expert}}}{\text{BW}_{\text{H2D}}}$,
where $M_{\text{expert}}$ is the expert weight size and $\text{BW}_{\text{H2D}}$ is the achievable PCIe bandwidth under contention. On H100-servers with PCIe Gen5 x16, we achieve $\text{BW}_{\text{H2D}} \approx 26$ GB/s per GPU.

\textbf{Contention Factor ($\eta$).} We observe an asymmetry in interference sensitivity: while contention degrades H2D transfer by 40\%, its impact on All2All is relatively minor, yielding $\eta \approx 0.19$. 
This asymmetry can be attributed to the underlying physical topology. Cross-node All2All is a multi-stage process that depends not only on PCIe but also on network fabric (e.g., InfiniBand) traversal. Consequently, the local latency penalty introduced by contention is amortized.

\subsubsection{Optimization Problem Definition}
We formulate the selection of virtual experts as a constrained optimization problem in Equation \ref{eq:opt}. The objective is to minimize the layer latency $T_{i}$. A critical constraint is that the H2D transfer time must be hidden within the execution time of the previous layer to prevent stalls. This yields an upper bound $V_{\max}$ on the number of virtual experts. Since the search space $[0, V_{\max}]$ is small (typically $V_{\max} < 3$), we employ a greedy traversal strategy at runtime to solve for $V^*$ efficiently without incurring solver overhead.
\begin{equation}
\small
\begin{aligned}
V^* = \argmin_{V \in \mathbb{Z}_{\ge 0}} \ T_{i}^*(L_i, V) \quad\text{s.t.} \quad  T_{\text{H2D}}(V) \le T_{i} \implies V \le V_{\max}
\end{aligned}
\label{eq:opt}
\end{equation}

\textbf{Generalizability.} The gain ($\gamma_1$, $\gamma_2$, $\Delta$) and PCIe contention ($\text{BW}_{\text{H2D}}$, $\eta$) parameters are inherently hardware- and model-specific, with numerical values calibrated based on the setup in Section \ref{sec:5.1}. Crucially, the allocator adapts to diverse setups with minimal effort and without any architectural modifications. 
For new models, load-balancing parameters $\alpha$, $\beta$, and $k$ are rapidly re-calibrated via offline forward-pass tracing in under 30 minutes. 
For new hardware platforms, contention parameters $\eta$ and $\text{BW}_{\text{H2D}}$ are seamlessly updated using microbenchmarks. A detailed evaluation of cross-model and cross-platform generalizability is provided in Appendix \ref{apx:generalizability} and Section \ref{sec:abl}.

\section{System Implementation} 
\label{sec:sys}

\subsection{vExpert Allocator }
We detail the allocator design from three perspectives: (1) group-wise budgeting strategy that expands the optimization space beyond single-layer constraints; (2) physical-to-virtual mapping algorithm; and (3) non-blocking pipeline integration that applies these decisions with negligible runtime overhead.

\textbf{Group-wise Budgeting.}
Per-layer vExpert budget is constrained by the overlap requirement:
H2D latency must be masked by the time of previous layer, typically restricting $V_{max} \approx 1$.
Our insight is that load distribution are heterogeneous across layers. Layers with high $L_i$ (e.g. $L_i>0.7$) gains little from vExpert, while imbalanced layers (e.g. $L_i<0.5$) benefit greatly as shown in \autoref{fig_delta}.

Therefore, we introduce \textbf{group-wise slots} that partitions $G$ consecutive layers into a single group. The constraint is relaxed from a per-layer limit to a group-wise budget: $\sum_{i=1}^{G} V_i \le V_{\max}^* = G \cdot V_{\max}$,
$V_i$ is the number of virtual experts for layer $i$. It masks H2D latency behind the computation time of $G$ layers, allowing well-balanced layers to donate their slots to imbalanced layers and expanding the solution space.
This strategy requires a GPU buffer size of $G \cdot V_{\max} \cdot M_{\text{expert}}$. 
The choice of $G$ represents a trade-off between optimization space and memory overhead. 
In practice, we set $G=2$ to achieve load balance with architectural simplicity and modest overhead (see Appendix \ref{apx:group}).

\textbf{Physical-to-Virtual Mapper.}
We develop a three-step mapping with details provided in Appendix \ref{apx:p2y_map}.

\textit{Step 1: Number Assignment.} To determine the optimal budget,  we first evaluate the $L_i$ of each layer within a group based on historical statistics.
Well-balanced layers ($L_i > 0.7$) are assigned $V_i = 0$, and their slots are redistributed to imbalanced layers, whose $V_i$ are generated via solving Equation \ref{eq:opt}. 

\textit{Step 2: Candidate Selection.} 
To identify the expert candidates for the assigned budget, for each layer $i$, we select the top $V_i \times N_{gpu}$ hottest physical experts from host DRAM based on historical statistics.

\textit{Step 3: Greedy Rebalance.} To ensure load balance across devices, we employ a greedy bin-packing approach to map candidates to physical GPUs: (i) Rank GPUs by their current token loads;
(ii) Iteratively assign the virtual expert with the highest pending load to the GPU with the lowest current load;
(iii) Repeat until all $V_i$ slots for the layer are filled.

\textbf{Pipeline Integration.}
To minimize interference with critical path, we integrate vExpert allocation into inference pipeline as:

\textit{Lightweight vExpert Allocation.} 
The system triggers allocation every $k$ iterations. The Allocator consumes the latest activation statistics and determines the optimal mapping. We implement the allocation using efficient CUDA kernels, enabling the entire allocation to complete within 10 ms.

\textit{Non-Blocking Graph Update.} When new vExpert info is ready, the system updates CUDA graph by modifying the \texttt{SrcPtr} field of \texttt{MemCpy} nodes through CUDA runtime API to avoid expensive recapturing. Since CUDA graphs support asynchronous updates, this process is non-blocking.

\textit{Synchronization.} Once current batch finishes, GPU switches to the updated graph and executes the next batch. 
This pipeline enables adaptation to dynamic workloads with near-zero overhead. Concretely, it introduces at a 10 ms overhead every k iterations; with $k\geq10$, the per-iteration overhead is below 1 ms, negligible in practice. Thus, we set $k=10$ in our default implementation.

\subsection{Coordinator–Worker Disaggregated Management}
Deploying tiered storage in multi-process environments faces two critical system-level problems: 

\textbf{Pinned memory overhead.} Achieving peak H2D bandwidth requires registering weights as pinned (page-locked) to enable DMA zero-copy transfers, which forces GPUs to maintain a page table mapping physical pages to accessible addresses. This incurs a non-trivial memory overhead. For instance, DeepSeek-V3.1 consumes $\sim$2GB VRAM per rank, totaling 16GB for an 8-GPU node.

\textbf{Shared memory limits. } To make CPU-resident experts visible across multiple processes, standard implementations rely on shared memory, which is typically capped at 50\% of physical RAM. This is insufficient for full redundancy of massive models, especially with NUMA-duplicating to avoid cross-socket penalties: DeepSeek-V3.1 requires 1.3TB while the usable shared memory is 900GB.

The root cause of these overheads is the enforcement of a uniform access model: every GPU process requires both read and write access —read for its own GEMM operations and write (initiating H2D) to populate HBM vExpert slots. 

\textbf{Disaggregated Architecture.}
We propose a \textbf{Coordinator–Worker} architecture to bypass these limits. The core principle is an IO-centralized coordinator to hold the responsibility for data provisioning (pinning memory, initiating H2D copies). The worker processes only require read access to their local GPU buffers for computation. This separation eliminates redundant host memory pinning and bypasses the need for OS shared memory.
As shown in \autoref{fig_overview}, we designate Rank 0 as the coordinator process and Ranks $1 - 7$ as workers within a node:

\textit{Coordinator:} The coordinator allocates and pins the expert pool in host memory. It uses CUDA IPC to import worker GPU buffers via \texttt{cudaIpcOpenMemHandle} and writes via \texttt{cudaMemcpyAsync}.

\textit{Worker:} Each worker allocates its local GPU ping-pong buffer for caching vExpert weights, exports handles via \texttt{cudaIpcGetMemHandle} and shares them with coordinator during initialization. It focuses exclusively on kernel execution once coordinator signals data readiness.

 A critical design choice is the number of pinned physical experts ($N_{p}$) maintained in the L2 storage. Increasing $N_{p}$ enhances redundancy capacity and improves load balance. However, it comes at the cost of higher DRAM consumption. NUMA locality further complicates this trade-off: duplicating experts across NUMA domains can mitigate cross-NUMA penalties but doubles the DRAM footprint. We provide a detailed analysis of this trade-off in Section~\ref{sec:mem_effi} and store a single copy of every expert in L2 storage without NUMA duplication as default.

\section{Evaluation}
\label{sec5}
\subsection{Evaluation Setup}
\label{sec:5.1}
\textbf{Hardware.}
Experiments are conducted on an 8-node cluster; each node features 8$\times$H100-NVLink GPUs, 1.8 TB of DDR5-4800 RAM, and 4$\times$400 Gbps InfiniBand interconnects. This setup represents a typical high-end production environment. Additional details are provided in Appendix \ref{apx:hardware}. 

\textbf{Workloads and Model.}
To emulate dynamic online serving and load drifts, we construct a synthetic workload from multiple publicly available datasets: ShareGPT \cite{sharegpt}, LongBench \cite{longbench} and LiveCodeBench \cite{livecodebench}. ShareGPT provides real-world conversations. LongBench is used to simulate long-context scenarios with lengthy system prompts. LiveCodeBench is employed for code-generation workloads.
Experiments are conducted using 671B DeepSeek-V3.1 \cite{deepseekv3} by default.

\textbf{Baseline.}
We use SGLang, a state-of-the-art open-source LLM serving system for large-scale cluster inference, as our primary baseline. All decode instances are configured following the best-practice strategies recommended in \cite{sglangdeploy} to ensure a strong baseline (refer to Appendix \ref{apx:sgl_setup} for details). 
\textsc{vExpert} is compared against:

(1) \textbf{B0 (Vanilla)}: Vanilla SGLang deployment without explicit load balancing; (2) \textbf{B1 (EPLB)}: SGLang with Expert-Parallelism Load Balancing \cite{eplb}, which rebalances placement periodically based on historical statistics; (3) \textbf{B2 (Redundant)}: EPLB augmented with one GPU-resident redundant expert per device, representing a typical HBM-constrained redundancy strategy; (4) \textbf{Offload}: Offloads one expert per device to CPUs, triggering H2D transfers of one expert per layer \cite{es-moe}.

The choice of rebalance frequency for B1 and B2 involves a trade-off between load balance and rebalance-blocking overhead. Given that EPLB incurs a stall of $\ge 1\text{s}$ per rebalance \cite{sglang}, we configure B1 and B2 with rebalance iterations of 500 and 200 as low and high-frequency. 

\begin{figure}[!t]
  \vskip 0.2in
  \begin{center}
\centerline{\includegraphics[width=1.0\textwidth,height=0.286\textwidth]{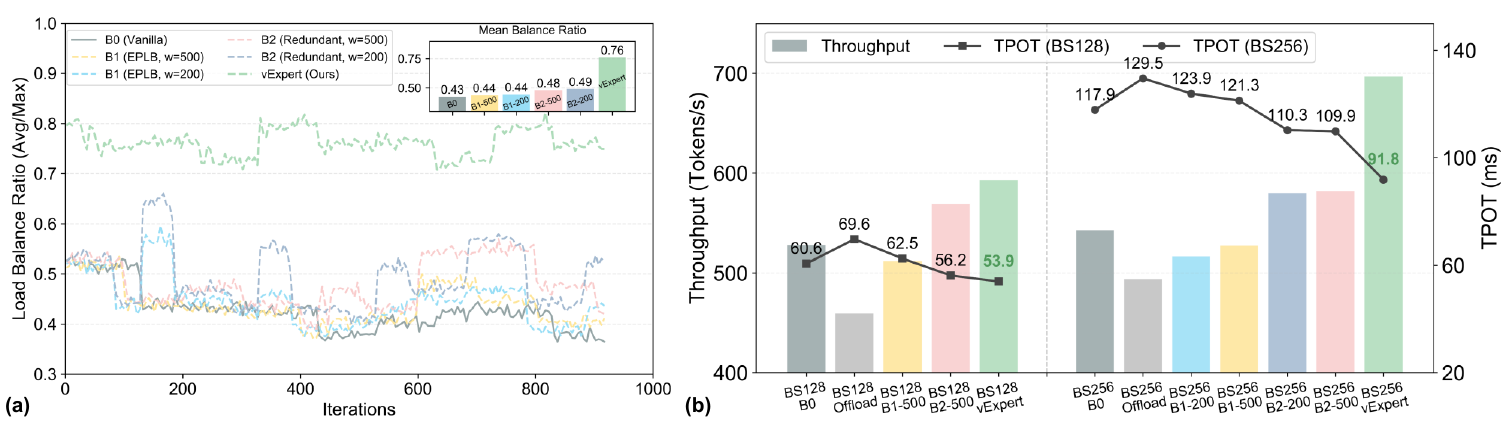}}
    \caption{(a) Comparison of load balance ratio over iterations at batchsize 256, w=$w$ denotes rebalancing every $w$ iterations.(b) End-to-end throughput (\textit{bars, left axis}) and Time-Per-Output-Token (TPOT) (\textit{lines, right axis}) across different batch sizes, B-$w$ denotes rebalancing every $w$ iterations.
    }
    \label{fig_exp}
  \end{center}
\vspace{-1.0cm}
\end{figure}

\subsection{Evaluation Results}
\textbf{Effectiveness in Load Balance.}
We evaluate \textsc{vExpert} in maintaining load balance under dynamic load drifts, using \textbf{load balance ratio} as metric. 
As shown in \autoref{fig_exp} (a), the vanilla baseline (B0) suffers from severe imbalance, with balance ratio at 0.43. While B1 and B2 provide marginal improvements (0.44 and 0.49), they are constrained by HBM capacity and the stalls induced by periodic rebalancing, failing to handle rapid load drifts.
In contrast, \textsc{vExpert} increases redundancy capacity and maintains load balance ratio at 0.76, improving by 55\% compared to the best baseline. 

\textbf{End-to-end Performance.}
\autoref{fig_exp} (b) highlights the end-to-end throughput gains. 
Due to constant H2D transfers, the penalty reduces throughput of Offload by 10\% compared to \text{B0}.
\text{B1} suffers degradation compared to \text{B0} since the overhead of rebalancing stalls outweigh the load-balancing benefits. \text{B2} achieves a modest 7\% gain via HBM-resident redundancy but faces diminishing returns at higher rebalancing frequencies, thus limiting its adaptability to dynamic workloads. 
By contrast, vExpert allocator enables seamless adaptation, translating to 28.4\% improvement at batch size 256 and 12.4\% at batch size 128 over \text{B0}.

The performance variance across batch sizes is attributed to H2D transfer overhead. Since $T_{\text{H2D}}$ depends only on the number of virtual experts $V$, its contribution to inference latency increases as batch size and TPOT decrease. Despite this, \textsc{vExpert} consistently outperforms all baselines, demonstrating its robustness in distributed environments.

\begin{figure}[ht]
    \centering
    \begin{minipage}[c]{0.60\textwidth}
        \centering
        \includegraphics[width=\textwidth]{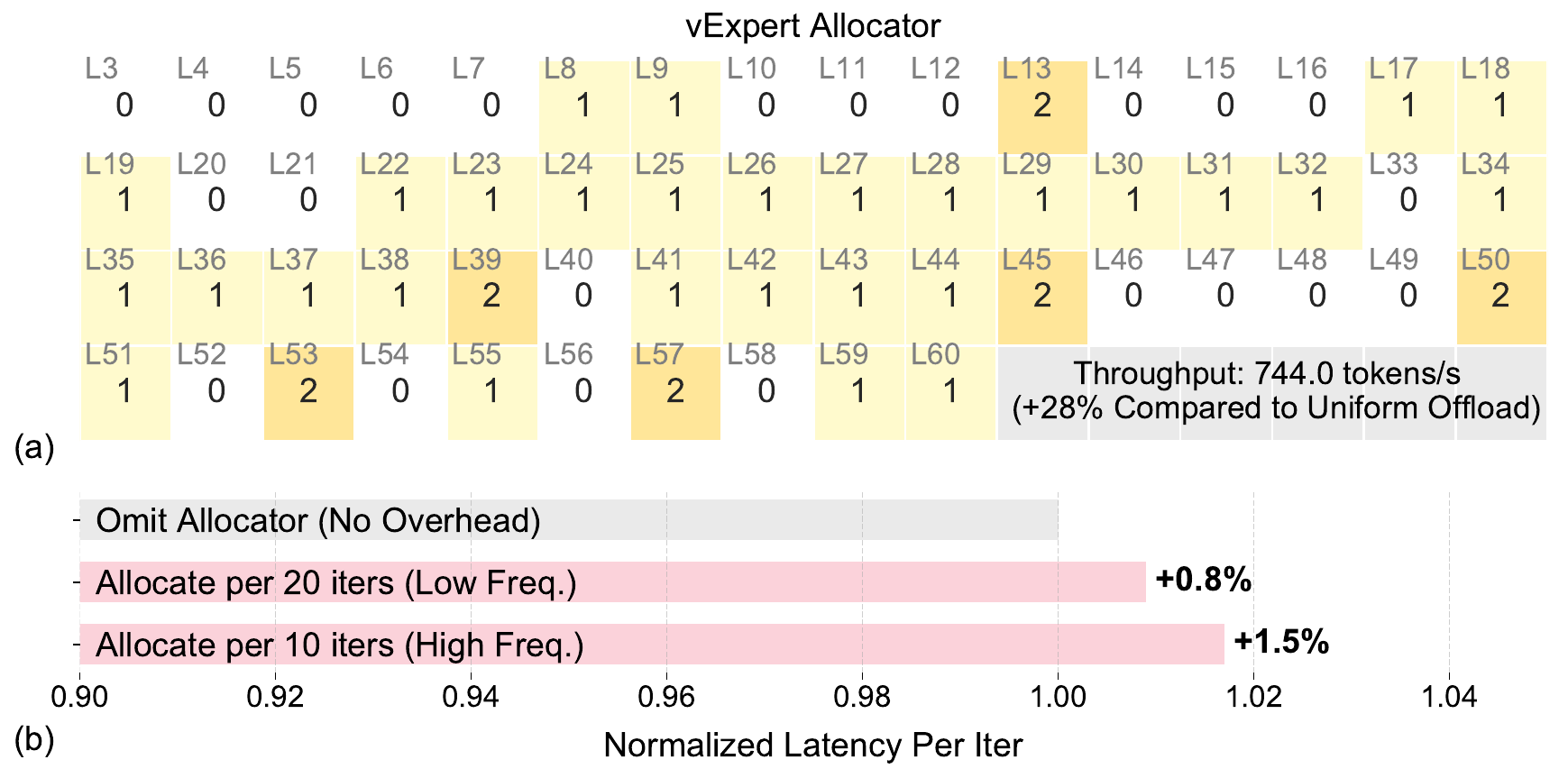}
    \end{minipage}\hfill
    \begin{minipage}[c]{0.37\linewidth}
        \centering
        \small
        \text{(c) Layer-wise allocation}\par
        \vspace{3pt}
        \setlength{\tabcolsep}{2pt}
        \renewcommand{\arraystretch}{1.1}
        \begin{tabular}{ccc}
            \toprule
            Slots per Layer & \# Layers & Before/After \\
            \midrule
            0 & 23 & 0.697/0.697 \\
            1 & 29 & 0.438/\textbf{0.763} \\
            2 & 6  & 0.347/\textbf{0.768} \\
            \bottomrule
        \end{tabular}
    \end{minipage}
    \caption{Allocator evaluation for DeepSeek-V3.1 on an 8-node H100 cluster at batch size 256.
    (a) Layer-wise assignment and throughput.
    (b) Allocator overhead.
    (c) Allocation across Layers 3--60.
    Before/After denotes the mean balance ratio before
    and after allocation.}
    \label{fig_allocator}
\vspace{-0.5cm}
\end{figure}

\subsection{Ablation Study}
\label{sec:abl}

\textbf{Efficiency of Allocator.}
Compared to uniformly offloading ($V_i=1$), our allocator yields a 28\% throughput gain by dynamically assigning virtual experts across layers, reflecting the layer-wise variation in load imbalance (\autoref{fig_allocator} (a)). To illustrate how the allocator adapts to this variation,
\autoref{fig_allocator} (c) summarizes the allocation across
Layers 3--60 in one DeepSeek-V3.1 iteration. The allocator assigns 41 virtual expert slots to 35 imbalanced
layers while leaving the other 23 relatively balanced layers
unchanged, avoiding unnecessary H2D transfers for these layers.
Layers receiving one and two slots improve their mean balance
ratios from 0.438 to 0.763 and from 0.347 to 0.768, respectively.
It demonstrates the effectiveness of adapting the number
of virtual experts to each layer's load imbalance.

We also evaluate the overhead of allocator as shown in \autoref{fig_allocator} (b). 
The penalty of enabling allocator is negligible (less than $2\%$). In addition, we provide detailed breakdown of allocator overhead in Appendix \ref{apx:allocator}.  While both 10-iter and 20-iter introduce minimal overhead, 10-iter provides superior adaptability to rapid load shifts. Consequently, we select 10-iter as default to ensure both agility and efficiency.

\begin{table*}[!t]
\centering
\caption{Effect of ablating memory manager on memory usage and system throughput measured at batchsize 256, ``N/A'' indicates a system crash or OOM due to memory constraints, ``NUMA'' denotes duplicating expert weight across two NUMA domains.}
\label{tab:memory_efficiency}
\small
\begin{tabular}{cccccc}
\toprule
\multirow{2}{*}{\text{Method}} & \multicolumn{1}{c}{\multirow{2}{*}{\begin{tabular}[c]{@{}c@{}}\text{Metric}\\ \text{(Per Node)}\end{tabular}}}  & \multicolumn{4}{c}{\text{Pinned Physical Expert Configuration}} \\
\cmidrule{3-6}
 &  &  \text{128} & \text{128-NUMA} & \text{256} & \text{256-NUMA} \\
\midrule
\multirow{3}{*}{\begin{tabular}[c]{@{}c@{}}Disable \\ Memory Manager\end{tabular}} 
& GPU Mem Alloc. & {\color{red!50!black}\text{6.9G}} & {\color{red!50!black}\text{10.4G}} & {\color{red!50!black}\text{11.9G}} & {\color{red!50!black}\text{18.4G}} \\
& CPU Mem Alloc./Avail. & 313G/{\color{red!50!black}\text{900G}} & 626G/{\color{red!50!black}\text{900G}} & 646G/{\color{red!50!black}\text{900G}} & 1.3T/{\color{red!50!black}\text{900G}} \\
  & Throughput & 1.00$\times$ & 1.02$\times$ & 1.05$\times$ & {\color{red!50!black}\textbf{N/A(OOM)}} \\ \midrule
\multirow{3}{*}{\begin{tabular}[c]{@{}c@{}}\text{Enable}\\ \text{Memory} \text{Manager}\end{tabular}} 
&  GPU Mem Alloc. & {\color{green!40!black}\text{0.9G}} & {\color{green!40!black}\text{1.3G}}  & {\color{green!40!black}\text{1.5G}} & {\color{green!40!black}\text{2.3G}}  \\
&  CPU Mem Alloc./Avail.& 313G/{\color{green!40!black}\text{1.8T}} & 626G/{\color{green!40!black}\text{1.8T}} & 646G/{\color{green!40!black}\text{1.8T}} & 1.3T/{\color{green!40!black}\text{1.8T}} \\
& Throughput &  1.00$\times$ & 1.02$\times$ & {\color{green!40!black}\textbf{1.05$\times$}}  & {\color{green!40!black}\textbf{1.06$\times$}} \\ \bottomrule
\end{tabular}
\vspace{-0.3cm}
\end{table*}

\textbf{Efficiency of Memory Manager.}
\label{sec:mem_effi}
We evaluate the manager on resource utilization and system throughput across different configurations. 

\textit{1. GPU Memory Savings.} As shown in Table~\ref{tab:memory_efficiency}, enabling the manager leads to an 8$\times$ reduction in GPU memory overhead. By centralizing management within the coordinator, \textsc{vExpert} avoids per-process pagetable-metadata overhead, with negligible HBM usage of less than 300MB per GPU. 

\textit{2. CPU Memory Utilization.} The manager enhances scalability by bypassing shared-memory limits. Without the manager, the system is constrained by the typical 50\% allocation of physical RAM for shared memory. \textsc{vExpert} utilizes the full DRAM capacity, allowing for higher redundancy capacity, which is critical for dynamic workloads.

\textit{3. Redundancy Capacity and NUMA Awareness.} Increasing the number of pinned experts improves throughput as a larger redundant expert pool increases the hit rate of hot experts in L2. Furthermore, replicating experts across NUMA domains also enhances performance.
Given that it doubles CPU memory usage, we select 256 as default with modest memory usage and throughput.

\textbf{Scalability with EP.}
Increasing the EP degree distributes expert ownership across
more ranks, making execution more sensitive to routing skew:
a rank owning a hot expert can become a more pronounced straggler
in A2A communication and MoE computation.
We conduct experiments by evaluating \textsc{vExpert}
at EP32 and EP64 on DeepSeek-V3.1 with batch size 256.
When scaling from EP32 to EP64, B2's balance ratio decreases
from 0.57 to 0.49, and its throughput drops by 11\%.
\textsc{vExpert} improves the balance ratio to 0.81 and 0.76
and increases throughput over B2 by 11\% and 20\%, respectively.
These results show that \textsc{vExpert} remains effective as EP
scales and expert ownership becomes more fragmented, with larger
relative gains under more severe load imbalance.

\textbf{Performance Model Robustness.} Table~\ref{tab:para} presents the sensitivity to $\pm20\%$ perturbations in each performance model parameter on DeepSeek-V3.1. Remarkably, even with substantial parameter errors, throughput degradation remains below $3\%$ in all cases. 
It demonstrates that \textsc{vExpert} does not require high-precision calibration to deliver near-optimal performance. The root cause is that the allocator’s primary decision variable—the number of virtual experts—is a \textbf{discrete integer} that naturally absorbs continuous perturbations. 

\begin{table}[t]
\centering
\begin{minipage}[t]{0.3\linewidth}
\centering
\caption{Throughput impact of parameter errors.}
\label{tab:para}

\begin{small}
\begin{tabular}{lcc}
\toprule
Para. & $-20\%$ Error & $+20\%$ Error \\
\midrule
$\alpha$ & $-1.2\%$ & $-2.7\%$ \\
$\beta$ & $-2.5\%$ & $-0.7\%$ \\
$k$ & $-2.4\%$ & $-2.8\%$ \\
$\text{BW}_{\text{H2D}}$ & $-0.9\%$ & $-1.4\%$ \\
$\eta$ & $-1.2\%$ & $-2.6\%$ \\
\bottomrule
\end{tabular}
\end{small}

\end{minipage}
\hfill  
\begin{minipage}[t]{0.6\linewidth}
\centering
\caption{Evaluation on KIMI-K2 (BatchSize 128, 4-Node H100 Cluster).}
\label{tab:kimi}
\begin{small}
\renewcommand{\arraystretch}{1.2}
\begin{tabular}{lcc}
\toprule
Method & Balance Ratio & Throughput(tok/s) \\
\midrule
B2-500 & $0.42$ & $545$ \\
Offload \cite{es-moe}   & $0.59$ & $529$ \\
\textbf{\textsc{vEXPERT}(Simplified)} & $0.64$ & $659$ ($\mathbf{+21\%}$) \\
\textbf{\textsc{vEXPERT}(Calibrated)} & $0.69$ & $703$ ($\mathbf{+29\%}$) \\
\bottomrule
\end{tabular}

\end{small}
\end{minipage}
\end{table}

\textbf{Cross-Model Generalizability.}
Table~\ref{tab:kimi} reports the results on different model KIMI-K2 \cite{kimi-k2}. We also implement a Simplified model that omits $\gamma_2$ and fixes the load-balance parameters $\alpha, \beta,k$ to constants derived from DeepSeek-V3.1 profiling.
The Simplified Model improves throughput by $\mathbf{+21\%}$. With full calibration, the gain rises to $\mathbf{+29\%}$. 
These results confirm that the structural insights captured by performance model are transferable across models. 
Results of Simplified Model further show that a coarse approximation of H2D impacts is sufficient to outperform offloading, underscoring that the joint modeling of PCIe contention, H2D latency, and load distribution is the critical enabler of robust performance.

\section{Conclusion and Limitation}
\label{sec6}
This work pioneeringly reveals and resolves challenges for exploiting host resource in distributed MoE inference.
By building systematic performance model and memory manager, it enables fine-grained and adaptive load balance, achieving up to 20\% throughput improvement and highly stable load balance under dynamic workloads. 
While \textsc{vExpert} targets the decode phase where host resources are underutilized, it complements GPU-based load-balancing methods (e.g., EPLB) in the prefill phase, as the longer per-forward latency in prefilling amortizes expert reordering overheads and host memory can be repurposed for prefix caching. 
Additionally, our analysis assumes conventional server topologies with PCIe/NVLink hierarchies. Emerging supernode architectures \cite{cloudmatrix,nvl72} featuring advanced packaging and high-bandwidth interconnects may reshape the host–device landscape. {For example, GB200 NVL72 introduces a fundamentally different hierarchy: coherent NVLink-C2C connects Grace CPUs and Blackwell GPUs, while up to 72 GPUs form a rack-scale NVLink domain. This higher-bandwidth CPU–GPU path could reduce host-expert transfer overhead, making host-backed expert virtualization promising on such systems. Adapting vExpert to advancing topology is an important direction for our future work.}

\section*{Acknowledgments}
This work was supported by National Key Research and Development Program of China under Grant (2024YFB4405400).

\small
\bibliographystyle{plain}   
\bibliography{ref}

\newpage

\newpage
\appendix
\section{Experimental Details}
\subsection{Hardware Configurations}
\label{apx:hardware}

All experiments are conducted on a multi-node GPU cluster representative of modern production inference environments.
Table~\ref{tab:hardware_config} summarizes the hardware configuration of each node.

Each node is equipped with 8 NVIDIA H100 GPUs (80GB HBM each), interconnected via NVLink within the node to provide high-bandwidth device-to-device communication. Nodes are connected through 4×400 Gbps InfiniBand links.
On the host side, each node contains dual Intel Xeon Platinum 8468V CPUs with 96 physical cores in total, and 1.8 TB of DDR5-4800 host memory split evenly across two NUMA domains. All GPUs are connected to the host via PCIe Gen5 ×16 links, yielding a peak per-GPU H2D bandwidth of up to 64 GB/s.
\subsection{Deployment Setup}
\label{apx:sgl_setup}
We deploy all methods using SGLang in disaggregated decode mode (version v0.5.1), following best-practice configurations for large-scale MoE inference \cite{sglangdeploy}.
All experiments focus exclusively on the decode stage, where strict latency constraints and dynamic expert activation patterns make load balancing particularly challenging.

\textbf{Decode Cluster Configuration.}
Each node runs 8 GPU worker processes, corresponding to one process per GPU. Decode workers across nodes are organized using data parallelism attention to maximize throughput while preserving expert parallelism semantics with parallel size equaling to total number of decode GPUs.

We enable \textbf{DeepEP} as the All-to-All backend with the low-latency mode and \textbf{DeepGEMM} as MoE operation backend. Two-batch overlap is enabled to overlap attention computation with MoE communication, following the execution timeline described in Section~3.
To improve batchsize and throughput of decode cluster while meeting service-level objective, all configurations are deployed with Multi-Token Prediction (MTP) speculative decoding enabled, configured with \textit{speculative-num-draft-tokens=4} with averaged accept length at $3.3$ \cite{h20deploy}.

We generate workload by injecting datasets into a standard Prefill-Decode disaggregattion pipeline. We sample requests from ShareGPT, LongBench, and LiveCodeBench. Injecting them sequentially to Prefill instances, simulating intra- and inter-dataset shifts. Decode instance fetches requests from Prefill instances to meet its target batch size. Following SGLang's stress-test protocal, we initially pause Decode instance to build a Prefill backlog, ensuring Decode queue remains fully saturated. 
\begin{table}[h]
\centering
\caption{Detailed Hardware Specifications per Computing Node.}
\label{tab:hardware_config}
\vspace{-0.5cm}
\vskip 0.15in
\begin{small}
\begin{tabular}{ll}
\toprule
\textbf{Component} & \textbf{Specification} \\
\midrule
CPU & Dual Intel Xeon Platinum 8468V (96 Physical Cores) \\
GPU & 8 $\times$ NVIDIA H100 (80GB HBM3) \\
Host Memory & 1.8 TB DDR5-4800  \\
Intra-node Interconnect & NVLink 4.0 \\
Inter-node Interconnect & 4 $\times$ 400 Gbps InfiniBand  \\
PCIe Interface & PCIe Gen5 $\times$ 16  \\
Peak H2D Bandwidth & 64$\times$8 GB/s \\
\bottomrule
\end{tabular}
\end{small}
\vspace{-0.5cm}
\end{table}

\section{Performance Model: Characterization, Calibration and Generalizability}
\label{apx:profiling}

In Section~\ref{sec:3.2}, we introduced a performance model to guide the vExpert allocator. 
This appendix details (1) the micro-benchmarking protocols used to calibrate hardware- and model-specific parameters ($\gamma_1, \gamma_2, \Delta, \eta$), 
(2) a lightweight calibration protocol for new deployment environments and (3) a rigorous evaluation of the model's robustness and generalizability across diverse models and hardware platforms.

\subsection{Experimental Setup}
Profiling is conducted on an 8-node H100 cluster as detailed in Appendix \ref{apx:hardware}. We use \texttt{torch.profiler} to capture CUDA graph execution timelines and kernel-level latencies. Unless otherwise specified, each H2D transfer in PCIe contention tests (\ref{apx:profiling_pcie}) moves a payload of $7168 \times 2048 \times 3 \times 2$ bytes per rank, corresponding to the weights of two experts in DeepSeek-V3.1.

\subsection{Characterizing Load Balancing Gains ($\gamma_1, \gamma_2, \Delta$)}
\label{apx:profiling_balance}

\begin{figure}[ht]
  \vskip 0.2in
  \begin{center}
    \centerline{\includegraphics[width=0.5\textwidth]{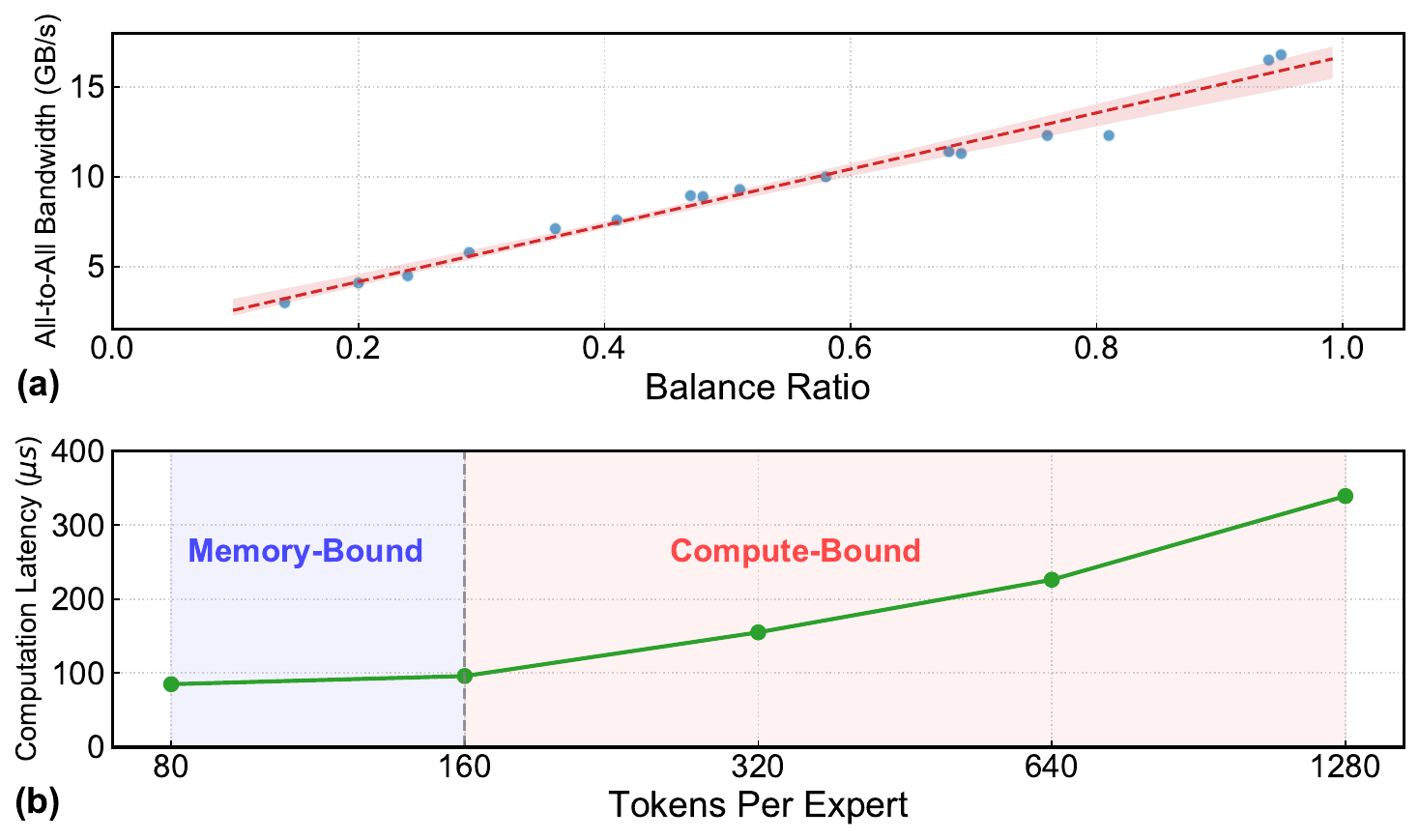}}
    \caption{
    \textbf{Impact of load balance ratio on All-to-All communication and MoE computation.}
    (a) Effective All-to-All bandwidth, measured as total data volume divided by latency.
    (b) MoE Kernel latency across different Tokens Per Expert (TPE) settings.
    }
    \label{fig_moe_a2a}
  \end{center}
  \vskip -0.2in
\end{figure}

We quantify how the load balance ratio $L_i$ impacts the latency of the critical path operations: All-to-All communication and MoE computation. We artificially skew expert assignments to vary $L_i$ and measure the resulting execution metrics using DeepEP and DeepGEMM backends.

\paragraph{Impact on All-to-All Communication ($\gamma_1$).}
Figure~\ref{fig_moe_a2a}(a) reports the effective All-to-All bandwidth. As load imbalance increases (lower Balance Ratio), the effective bandwidth drops sharply, even though the aggregate data volume remains constant. This behavior stems from the "straggler effect": the latency is dominated by the rank transmitting the maximum data volume. GPUs hosting “hot” experts saturate their links while lightly loaded GPUs idle early. Consequently, improving the balance ratio directly reduces the maximum per-rank token load, yielding a near-linear reduction in communication latency. This validates our approximation in Equation \ref{eq:gamma1}, where $\Delta(L_i, V)$ represents the improvement in load balance.

\paragraph{Impact on MoE Computation ($\gamma_2$).}
Variations in load balance result in different \textit{Tokens Per Expert} (TPE) assignments, which significantly affect MoE GEMM kernel performance. Figure~\ref{fig_moe_a2a}(b) illustrates two distinct regimes:
\textbf{1. Compute-Bound Regime (TPE $\geq 160$):} Kernel execution scales proportionally with the maximum token count. Reducing the straggler load yields substantial gains, with $\gamma_2 \approx \gamma_1$.
\textbf{2. Memory-Bound Regime (TPE $< 160$):} Execution is dominated by kernel launch overheads and memory access patterns. Reducing the maximum token load provides diminishing returns; latency remains largely flat despite TPE improvements, leading to $\gamma_2 \approx 0$.

\paragraph{Modeling Load Balance Improvement ($\Delta$).}
The improvement term $\Delta(L_i, V)$ is a complex function of the activation distribution, the initial imbalance $L_i$, and the redundancy budget $V$. Deriving a closed-form solution is intractable due to stochastic online request distributions. We instead characterize this relationship through offline profiling on representative datasets (ShareGPT, LongBench). Our observations indicate that for a fixed $V$, $\Delta$ approximates a linear function of the initial balance ratio. We therefore model $\Delta$ via a first-order linear approximation:
\begin{equation}
\Delta(L_i, V) = \alpha \cdot L_i + \beta \cdot V + k
\end{equation}

\subsection{Characterizing PCIe Contention ($\eta$ and $\text{BW}_{H2D}$)}
\label{apx:profiling_pcie}

\begin{figure}[ht]
  \vskip 0.2in
  \begin{center}
    \centerline{\includegraphics[width=1.0\textwidth]{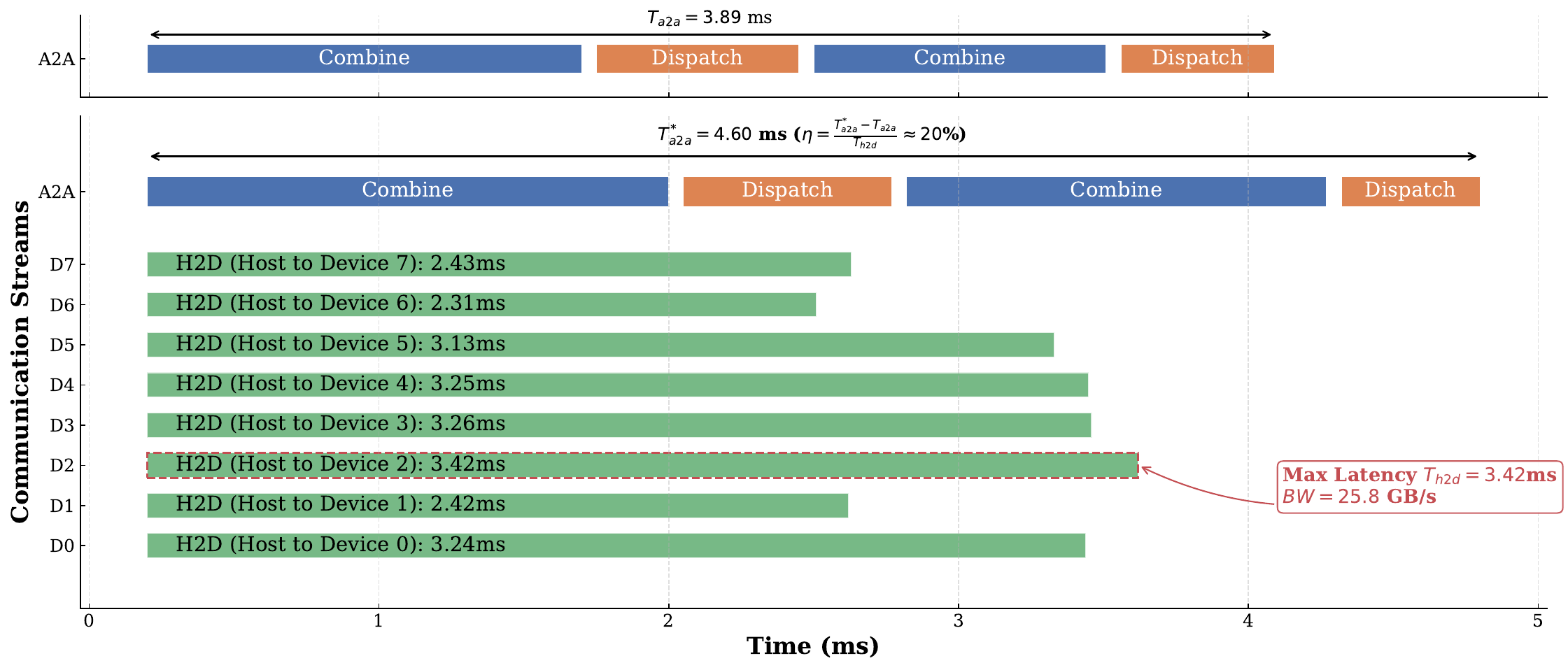}}
    \caption{
    \textbf{Micro-benchmark of PCIe contention on H100 cluster (Batch Size 128).}
    \textbf{Top:} Baseline execution timeline (A2A only) showing stable latencies.
    \textbf{Bottom:} Concurrent execution (A2A + H2D). The overlap introduces resource contention, leading to heterogeneous H2D throughput and increased latency for both operations.
    }
    \label{fig_pcie_detail}
  \end{center}
  \vskip -0.2in
\end{figure}

This section quantifies the hardware-level interference between Host-to-Device (H2D) expert streaming and cross-node All-to-All (A2A) communication.

\paragraph{Timeline Analysis and Observations.}
\autoref{fig_pcie_detail} illustrates the execution timeline for both the baseline (A2A only) and concurrent (A2A + H2D) scenarios. In the baseline scenario, the \texttt{Combine} and \texttt{Dispatch} operators on A2A Stream exhibit stable latencies. In the concurrent scenario, H2D MemCpy operations across multiple streams reveal two key effects:

\textbf{1. Intra-node H2D Variation:} H2D completion times are non-uniform across devices within the same node. This variance stems from shared PCIe tree topologies, where concurrent DMA requests from multiple GPUs compete for the finite arbitration bandwidth of PCIe switches and the CPU's Root Complex.

\textbf{2. Cross-Operation Interference:} Both A2A and H2D transfers on auxiliary streams suffer from increased latency. Although PCIe is theoretically full-duplex, control signaling and aggregate data pressure on internal switch fabrics lead to measurable throughput degradation for both RDMA-based A2A and PCIe-based H2D.

\paragraph{Quantitative Modeling.}
We define the contention penalty $\eta$ based on the degradation of A2A latency:
\begin{equation}
\eta = \frac{T_{a2a}^* - T_{a2a}}{T_{h2d}}
\end{equation}
where $T_{a2a}$ is the baseline latency, $T_{a2a}^*$ is the latency under contention, and $T_{h2d}$ is the duration of the concurrent transfer. Our measurements yield $\eta \approx 20\%$ and an effective $\text{BW}_{H2D}$ that is degraded compared to the theoretical peak. These values are integrated into the vExpert Allocator (Eq. \ref{eq:opt}) to predict whether improved load balancing outweighs PCIe contention delays.

\subsection{Calibration Protocol for New Environments}
\label{apx:calibration}

While the parameter values presented above are specific, the performance modeling methodology (Section \ref{sec:3.2}) is hardware- and model-agnostic. A key advantage of VEXPERT is the minimal recalibration effort required for deployment on new model architectures or hardware platforms. We prescribe the following lightweight, two-path calibration protocol:

\textbf{1. Model-Specific Path ($\alpha, \beta, k$):} 
    For a new model architecture, run the target model's gating network on representative validation datasets to capture expert activation distribution. Vary the number of virtual experts $V$ and record the resulting load balance ratio $L_i$ for each layer. A linear regression on these data points extracts the model-specific coefficients $\alpha$, $\beta$, and $k$. This tracing completes in under 30 minutes.

\textbf{2. Hardware-Specific Path ($\text{BW}_{H2D}, \eta$):}
    For a new hardware cluster, only the PCIe contention parameters require re-characterization. Run the concurrent micro-benchmark described in Section \ref{apx:profiling_pcie} to measure the effective H2D bandwidth ($\text{BW}_{H2D}$) and calculate $\eta$ using Equation 8. This single-step procedure captures the interference patterns of the deployment cluster’s PCIe switches and CPU Root Complexes, completing in under 10 minutes.

New models only require Step~1 (tracing $\alpha, \beta, k$), while new hardware only requires Step~2 (measuring $\text{BW}_{\text{H2D}}, \eta$). Once calibrated, parameters are loaded as a static configuration, ensuring the runtime optimization reflects the physical constraints.

\subsection{Generalizability Evaluation}
\label{apx:generalizability}

A critical concern for performance model is whether its benefits persist under imperfect calibration or when deployed on unseen models and hardware. We systematically evaluate robustness along: cross-model validation, and hardware generalizability. The cross-model generalizability is evaluated at Section \ref{sec:abl}, here we discuss the cross-platform generalizability in detail.

\subsubsection{Hardware Generalizability}
\label{apx:hardware_generalizability}

While our primary evaluation targets NVIDIA H100 clusters, the vExpert performance model is architecturally portable. Modern servers---including NVIDIA H800/H20 and AMD MI300X---universally adopt \textbf{PCIe Gen5} interconnects. The consistent physical topology (shared PCIe switch trees, CPU Root Complex arbitration) dictates that \textbf{PCIe contention behavior remains structurally similar}, and parameters $BW_{\text{H2D}}$ and $\eta$ remain stable across platforms with comparable PCIe generations.

For older PCIe Gen4 hardware, we provide a quantitative sensitivity analysis showing that vExpert's benefits are not merely preserved but in fact \textit{amplified}. From Section~\ref{sec:3.2}, the normalized H2D overhead is:
\begin{equation}
O = \frac{T_{\text{H2D}} \cdot \eta}{T_{\text{layer}}} = \eta \cdot \frac{V_{m}}{\text{BW}_{\text{H2D}} \cdot T_{\text{layer}}}
\end{equation}
where $V_{m}$ is the per-expert transfer size, $BW_{\text{H2D}}$ is the PCIe bandwidth, and $T_{\text{layer}}$ is the layer computation time. Halving the PCIe bandwidth would, in isolation, double the overhead. However, \textbf{PCIe bandwidth and compute capability scale together} across server generations:
\begin{itemize}[leftmargin=*]
    \item \textbf{H100 Server:} PCIe Gen5 ($\approx$64 GB/s) with 1,979 TFLOPS (FP16).
    \item \textbf{A100 Server:} PCIe Gen4 ($\approx$32 GB/s) with 312 TFLOPS (FP16).
\end{itemize}

Because $T_{\text{layer}}$ is approximately inversely proportional to peak FLOPS, the relative overhead on Gen4 compared to Gen5 is:
\begin{equation}
\frac{O_{\text{Gen4}}}{O_{\text{Gen5}}} \approx \frac{\text{BW}_{\text{Gen5}}}{\text{BW}_{\text{Gen4}}} \cdot \frac{T_{\text{layer}}^{\text{(Gen5)}}}{T_{\text{layer}}^{\text{(Gen4)}}} \approx \frac{64}{32} \cdot \frac{312}{1979} \approx 0.32
\end{equation}

Thus, VEXPERT's gains are preserved on lower-bandwidth platforms since slower compute provides more overlap headroom for H2D transfers.

\section{End-to-end Timeline Comparison}
\label{apx:timeline}
\begin{figure}[ht]
\vspace{-0.6cm}
  \vskip 0.2in
  \begin{center}
    \centerline{\includegraphics[width=1.0\textwidth,height=0.31\textwidth]{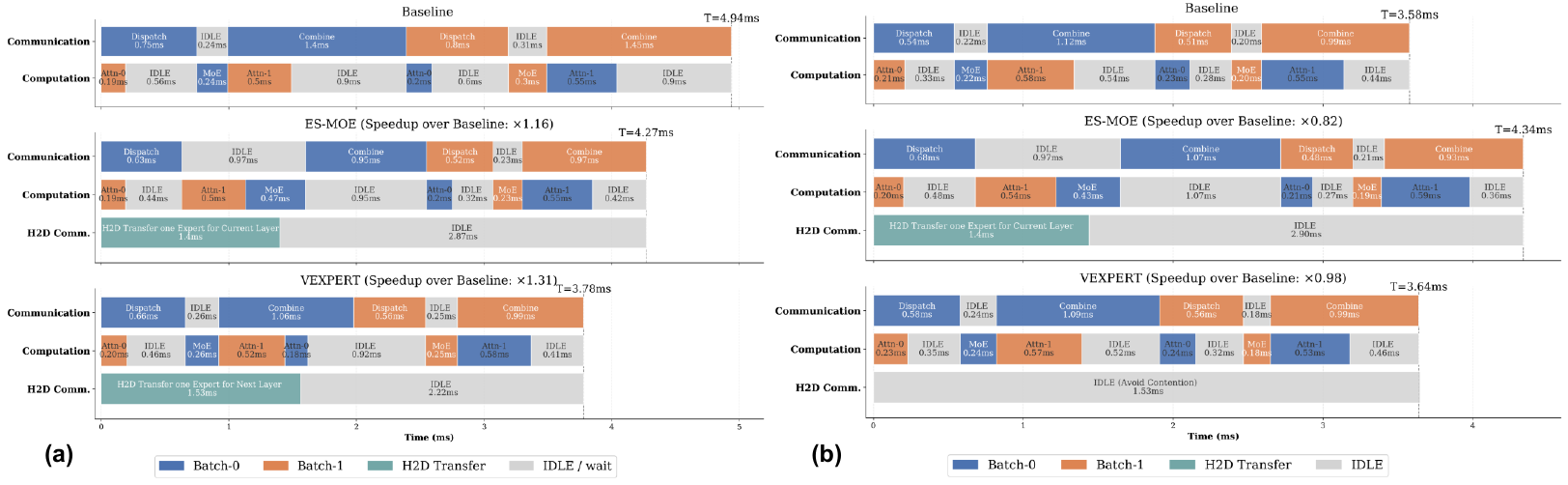}}
\caption{
    End-to-end timeline comparison on DeepSeek-V3.1 using a 4-node H100 cluster with a batchsize of 256. (a) Timeline of an imbalanced layer with a balance ratio of 0.45. (b) Timeline of a balanced layer with a balance ratio of 0.68.
  }
    \label{fig_v1_timeline_comp}
  \end{center}
\vspace{-0.6cm}
\end{figure}
Figure~\ref{fig_v1_timeline_comp} presents a detailed per-layer timeline comparison among \textbf{vExpert}, \textbf{Baseline} (B2, as described in Section~\ref{sec:5.1}), and \textbf{ES-MoE} \cite{es-moe} (offloading-based load balancing, labeled as ``Offload'' in Section~\ref{sec:5.1}). The traces are captured on a 4-node H100 cluster running DeepSeek-V3.1 at batch size 256. We select two representative layers to illustrate the spatial dynamism of load distribution.

\paragraph{Limitations of Baseline (B2).}
Due to the inherent \textbf{spatial dynamism} (layer-to-layer variation) and \textbf{temporal dynamism} (iteration-toiteration fluctuation) of expert activation distributions, the Baseline struggles to achieve global load balance. While static expert placement may yield acceptable balance for some layers, it inevitably leaves others severely skewed. 

\paragraph{Limitations of Offloading (ES-MoE).}
ES-MoE leverages host memory to re-balance tokens across devices via offloading partial experts. However, this approach suffers from two fundamental shortcomings. 
\textbf{First}, it ignores the \textbf{PCIe contention} introduced by H2D transfers. Streaming experts via PCIe shares the interconnect fabric with RDMA-based A2A communication. 
\textbf{Second}, ES-MoE applies a \textit{uniform} offloading policy across \textit{all} layers, unable to exploit inter-layer and inter-inter heterogeneity in load distribution. Even for naturally balanced layers (\autoref{fig_v1_timeline_comp} (b)), it incurs \textbf{constant H2D transfers} that introduce unnecessary synchronization overhead and contention penalties.
Consequently:

\textbf{For Imbalanced Layer:} VEXPERT utilizes H2D to prevent load imbalance that blocks All-to-All, yielding to a 30\% latency reduction.

\textbf{For Balanced Layer:} VEXPERT skips unnecessary H2D to avoid contention while offloading requires constant transfers.

\section{Analysis on vExpert Group}
\label{apx:group}
\begin{figure}[ht]
  \vskip 0.2in
  \begin{center}
    \centerline{\includegraphics[width=1.0\textwidth,height=0.4\textwidth]{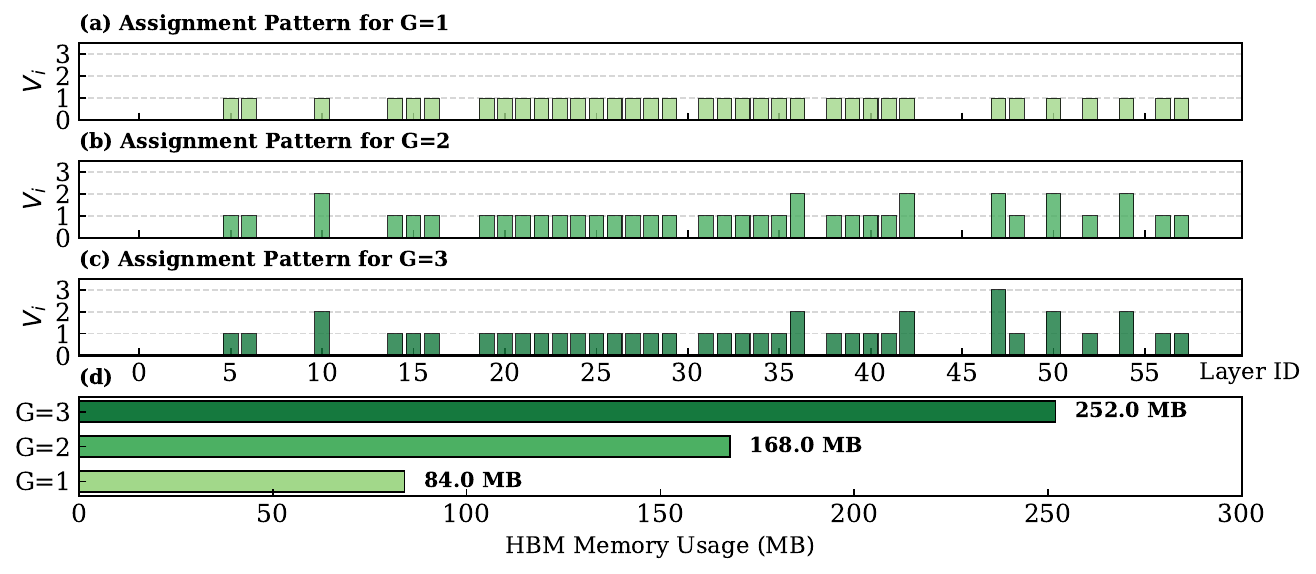}}
\caption{
    \textbf{Analysis of vExpert group size $G$ on assignment pattern and memory overhead.} 
    (a)-(c) illustrate the layer-wise assignment of virtual experts ($V_i$) across different group sizes. 
    (d) The GPU HBM overhead for required slot buffers scales linearly with $G$.
  }
    \label{fig_group}
  \end{center}
\vspace{-0.6cm}
\end{figure}
In this section, we analyze the impact of group size $G$. The group-wise slot strategy is designed to relax the strict per-layer H2D bandwidth constraint, allowing the system to redistribute vExpert slots from "easy-to-balance" layers to "bottleneck" layers.

\paragraph{Trade-off between Flexibility and Memory.} The total GPU memory overhead for vExpert slots is defined as $M = G \cdot V_{\max} \cdot W_{\text{expert}}$. While a larger $G$ expands the search space for the optimization problem (Eq. \ref{eq:opt}), it also increases the HBM footprint. We evaluate $G \in \{1, 2, 3\}$ as shown in \autoref{fig_group}. 

$G=1$ (Rigid Constraint): Each layer is strictly limited by its own TPOT. If a layer is highly imbalanced but its TPOT is short, it cannot acquire more than $V_{\max}$ experts, even if adjacent layers have significant idle PCIe bandwidth.

$G=2$ (Elastic Allocation): By grouping two layers, the allocator can assign $V_i=2$ to an imbalanced layer by setting $V_{i+1}=0$ for its balanced neighbor. This slot donation mechanism improves the load-balancing effectiveness for critical layers.

$G=3$ (Diminishing Returns): While $G=3$ further relaxes the constraint, our experiments show that the assignment pattern remains largely similar to $G=2$. Most imbalanced layers are effectively covered by 2 experts, and the additional memory cost for $G=3$ does not translate into higher throughput.

As shown in our evaluation, the layer-wise assignment for $G=2$ differs from $G=1$, specifically targeting imbalanced layers. In contrast, the assignment for $G=3$ provides marginal gains over $G=2$. 
Beyond hardware resource consumption, we consider the system orchestration overhead as a decisive factor: 

\textbf{Implementation Complexity:} A larger $G$ necessitates a more sophisticated scheduler to manage multi-layer dependencies and a more complex memory manager to track asynchronous H2D transfers across an extended window. Such complexity risks bloating the runtime logic and increasing the host-side CPU overhead. Maintaining a small $G$ ensures that the vExpert allocator remains lightweight and agile, allowing it to complete reconfigurations within the strict latency constraints of real-time decoding. Given that $G=2$ achieves the optimal balance between balancing flexibility and implementation simplicity—while only incurring a modest HBM footprint of $\sim$168 MB (a negligible fraction of the 80GB HBM on an H100 GPU)—we select $G=2$ as the default configuration for deployment.

\begin{algorithm}[t]
\caption{Physical-to-Virtual Rebalancing}
\label{alg:p2v_mapping}
\begin{algorithmic}[1]

\REQUIRE
Expert load statistics $W$;
group size $G$;
per-layer vExpert budget $V_{\max}$;
number of GPUs $N_{\text{gpu}}$;
Inference constant $C$;

\ENSURE
Physical-to-virtual map $\Phi$;

\STATE \textbf{// Step 1: Group-wise vExpert number assignment}
\FOR{each layer group $\mathcal{G}$ of size $G$}
    \STATE Compute group budget $V_{\max}^{*} \leftarrow G \cdot V_{\max}$
    \FOR{each layer $L_i \in \mathcal{G}$}
        \STATE Compute GPU-wise load $\mathbf{W}_i$
        \STATE $\mathcal{L}_i \leftarrow \mathrm{mean}(\mathbf{W}_i) / \max(\mathbf{W}_i)$
    \ENDFOR
    \STATE $\{V_i\}_{L_i \in \mathcal{G}} \leftarrow$
    Algorithm~\ref{alg:allocate_vexpert}$(\{\mathbf{W}_i, \mathcal{L}_i\}_{L_i \in \mathcal{G}}, V_{\max}^{*}, C)$  \text{// vExpert Number Assignment}
\ENDFOR

\STATE \textbf{// Step 2: vExpert Candidate selection}
\FOR{each layer $L_i$}
    \STATE Rank physical experts by activation frequency
    \STATE Select top $V_i \times N_{\text{gpu}}$ physical experts
    \STATE Generate vExpert-to-physical Candidate $\Psi$
\ENDFOR

\STATE \textbf{// Step 3: Greedy physical rebalancing}
\FOR{each layer $L_i$}
    \STATE Initialize GPU loads with base experts
    \FOR{each virtual expert $v$ in descending load order of $\Psi$}
        \STATE Select GPU $g$ with minimum current load
        \STATE Assign expert $v$ with the largest load to GPU $g$
        \STATE Update GPU load and capacity
        \STATE Update mappings $\Phi$
    \ENDFOR
\ENDFOR

\STATE \textbf{return} $\Phi$

\end{algorithmic}
\end{algorithm}

\begin{algorithm}[t]
\caption{Group-wise vExpert Number Assignment}
\label{alg:allocate_vexpert}
\begin{algorithmic}[1]

\REQUIRE
Layer-wise GPU load $\{\mathbf{W}_i\}$;
layer balanceness $\{\mathcal{L}_i\}$;
group vExpert budget $V_{\max}^{*}$;
inference constant $C$

\ENSURE
vExpert count per layer $\{V_i\}$

\STATE
Initialize $V_i^{\max} \leftarrow V_{\max}^{*} / |\mathcal{G}|$ for all $L_i \in \mathcal{G}$
\STATE \textbf{// Filter well-balanced layers }
\FOR{each layer $L_i \in \mathcal{G}$}
    \IF{$\mathcal{L}_i > \tau$}
        \STATE $V_i^{\max} \leftarrow 0$
    \ENDIF
\ENDFOR
\STATE \textbf{// Redistribute slots within the group}
\FOR{each layer $L_i \in \mathcal{G}$}
    \STATE Adjust $\{V_i\}$ to satisfy $\sum_{L_i \in \mathcal{G}} V_i = V_{\max}^{*}$
\ENDFOR

\STATE \textbf{// Optimization-based redistribution}
\STATE
Initialize $V_i \leftarrow V_{\max}^{*} / |\mathcal{G}|$ for all $L_i \in \mathcal{G}$
\FOR{each remaining layer $L_i \in \mathcal{G}$}
    \STATE $t_{\min} \leftarrow +\infty$
    \STATE $V_i^{\star} \leftarrow 0$
    \FOR{$v = 0$ to $V_i^{\max}$}
        \STATE Estimate communication time $t_{\text{comm}}(v; C)$
        \STATE Estimate MoE compute time $t_{\text{moe}}(v; C)$
        \STATE Estimate H2D transfer overhead $t_{\text{H2D}}(v,C)$
        \STATE Compute objective $t(v)$ using Eq.~(\ref{eq:layer_time_model})
        \IF{$t_i(v) < t_{\min}$}
            \STATE $t_{\min} \leftarrow t_i(v)$
            \STATE $V_i^{\star} \leftarrow v$
        \ENDIF
    \ENDFOR
    \STATE $V_i \leftarrow V_i^{\star}$
\ENDFOR

\STATE \textbf{return} $\{V_i\}$

\end{algorithmic}
\end{algorithm}

\section{Physical-to-virtual Mapping Design}
\label{apx:p2y_map}
This appendix details the proposed physical-to-virtual mapping strategy.
The rebalancing procedure is summarized in Algorithm~\ref{alg:p2v_mapping} and
consists of three stages: group-wise vExpert number assignment, vExpert selection,
and greedy physical rebalancing.

\subsection{Step 1: Group-wise vExpert Number Assignment}

Layers are partitioned into groups of size $G$.
Each layer is initially allocated a per-layer vExpert budget $V_{\max}$ which is determined by TPOT SLO.
Within a group $\mathcal{G}$, the total vExpert budget is therefore
$V_{\max}^{*} = G \cdot V_{\max}$.

For each layer $L_i \in \mathcal{G}$, we compute the layer balance ratio $\mathcal{L}_i$
 using expert load statistics $\mathbf{W}_i$ from previous iterations: 
\begin{equation}
\mathcal{L}_i = \frac{\mathbb{E}[\mathbf{W}_i]}{\max(\mathbf{W}_i)} .
\end{equation}

Well-balanced layers are filtered to assign zero slot, allowing their
unused budget to be redistributed to more imbalanced layers within the same group.
The final vExpert count $\{V_i\}_{L_i \in \mathcal{G}}$ is determined by
Algorithm~\ref{alg:allocate_vexpert}, which minimizes an estimated per-layer
iteration cost based on Eq. \ref{eq:opt}.
The optimization is parameterized by an inference constant $C$, which captures
runtime-dependent factors such as batch size and hardware communication
characteristics.

\subsection{Step 2: vExpert Candidate Selection}

Given the vExpert count $V_i$ for each layer, we select physical experts to
replicate.
Physical experts are ranked by their historical activation frequency.
For layer $L_i$, the top $V_i \times N_{\text{gpu}}$ experts are selected and
replicated to form virtual experts.
This produces a vExpert-to-physical mapping candidates that is reused in the subsequent
rebalancing stage.

\subsection{Step 3: Greedy Physical Rebalancing}

Virtual experts are assigned to GPUs using a greedy bin-packing strategy.
Each GPU is treated as a bin whose capacity equals the number of base experts plus
the allocated vExperts.
For each layer, virtual experts are processed in descending order of their expected
load.
Each virtual expert is assigned to the GPU with the smallest current load that still
has available capacity.
After each assignment, GPU load statistics and expert mappings are updated
accordingly.
This greedy strategy is efficient and empirically achieves near-optimal load balance.

\section{Analysis of Allocator Overhead}
\label{apx:allocator}
In this section, we provide a detailed breakdown of the latency overhead introduced by the vExpert Allocator. To maintain high throughput in real-time decoding, it is imperative that the reconfiguration logic does not become a bottleneck. We evaluate the overhead on our cluster consisting of 8 computing nodes (64 GPUs total).

The allocation process occurs every $k$ iterations and consists of two primary stages:

\textbf{1. Global Statistics Aggregation (4 ms):} Each rank maintains a local \texttt{logical\_count} tensor, recording the activation frequency of experts. Before allocation start, system performs \texttt{All-Reduce} operation across all GPUs to synchronize these statistics. Despite the cluster scale, this collective communication is highly optimized via InfiniBand, completing in approximately \textbf{4 ms}.

\textbf{2. Physical-to-Virtual Mapping (7 ms):} Once the global activation distribution is available, each rank executes the physical-to-virtual mapping (as detailed in Appendix \ref{apx:p2y_map}). This computational step consumes only \textbf{7 ms}.

The total latency for a single reconfiguration event is:
\begin{equation}
    T_{\text{alloc}} = T_{\text{all\_reduce}} + T_{\text{mapping}} = 11 \text{ms}
\end{equation}

In our default configuration, we set the allocation frequency at $k=10$. This allows the system to remain responsive to load drifts while significantly diluting the overhead at $\frac{T_{\text{alloc}}}{k} = 1.1\text{ms per iter}$. This confirms that vExpert is a lightweight and scalable solution suitable for high-frequency online adaptation.

\end{document}